# Evolution of magnetism in Pd-substituted $Ce_2RhIn_8$ single crystals

*M. Kratochvilova[1], K. Uhlirova[1], J. Prokleska[1], S. Danis[1], P. Cermak[1], V. Sechovsky[1]*

[1] *Department of Condensed Matter Physics, Charles University in Prague, Faculty of Mathematics and Physics, Ke Karlovu 5, 121 16 Prague 2, Czech Republic*

**Abstract:**

The evolution of magnetism and superconductivity in $Ce_2Rh_{1-x}Pd_xIn_8$ solid solutions has been studied within the entire concentration range by means of thermodynamic and magnetic measurements at ambient pressure and at temperatures between 0.35 K and room temperature. For this purpose, single crystals with Pd concentrations $x = 0$, 0.10, 0.15, 0.30, 0.45, 0.55, 0.85 and 1 have been grown from In self-flux and characterized by x-ray diffraction and microprobe analysis. Starting from the antiferromagnet $Ce_2RhIn_8$, the Néel temperature gradually decreases with increasing Pd concentration and the antiferromagnetism has disappeared for $x \geq 0.45$. Superconductivity has been observed only for $Ce_2PdIn_8$.

## 1. Introduction

$Ce_nTIn_{3n+2}$ ($n = 1, 2$; $T =$ Co, Ir, Rh) heavy-fermion (HF) compounds have been attracting attention of scientific community for almost two decades owing to coexistence of magnetic order and superconductivity in a wide area of the temperature-pressure-composition phase diagrams [1]. In the vicinity of the magnetic quantum critical point (QCP), unconventional superconductivity has been reported [2]. The compounds crystallize in the tetragonal $Ho_nCoGa_{3n+2}$-type of structure (P4/mmm). The layered character of the structure is determined by various sequences of $TIn_2$ and $CeIn_3$ basic building blocks stacked along the *c*-axis. The dimensionality of the structure, which spans from 3D ($CeIn_3$) to quasi-2D ($CeTIn_5$) seems to control the type of ground state. The desire to examine this aspect in more detail has motivated further efforts to synthesize and investigate new members of the $Ce_nTIn_{3n+2}$ family [3].

$Ce_2RhIn_8$ is a prototype example of a $Ce_nTIn_{3n+2}$ HF antiferromagnet exhibiting pressure induced superconductivity [2]. It orders below $T_N = 2.8$ K in a commensurate antiferromagnetic (AF) structure with (½, ½, 0) propagation vector and with the Ce magnetic moments making an angle of 52º with the basal plane and successively displays a second transition to the ground-state incommensurate magnetic structure at $T_{LN} = 1.65$ K [4,5]. Analysis of thermal-expansion, magnetovolume and specific-heat data indicates that the physical properties of $Ce_2RhIn_8$ result from competition between magnetic exchange, Kondo and crystal-field interactions[5]. Upon increasing the magnetic field along the *a*-axis, $T_N$ of $Ce_2RhIn_8$ increases and two magnetic-field-induced transitions appear. The phase transitions divide the phase diagram into several regions with different magnetic structures which have not been studied microscopically so far. $T_N$ decreases monotonically with increasing magnetic field applied along the *c*-axis. The ground state of $Ce_2RhIn_8$ can be tuned from AF to a superconducting (SC) not only by applying pressure [6] but also by substitutions [7,8]. Hydrostatic pressure induces a SC phase which coexists with the AF phase up to a pressure of 1.7 GPa and, at higher pressures, only the superconductivity remains [6]. Compared to its

more-2D analogue, the $Ce_2RhIn_5$ system, the effect of various substitutions in $Ce_2RhIn_8$ has been studied only poorly. Particularly interesting is the evolution of the critical temperatures in the case of transition-metal-site substitutions. Ir-rich samples of the $Ce_2Rh_{1-x}Ir_xIn_8$ series undergo a SC transition; another SC state in the low-Ir-concentration region can be induced by applying hydrostatic pressure [7]. The effect of Ir substitution exhibits similarities with the behavior of $CeRh_{1-x}Ir_xIn_5$ [9]. Co substitution for Rh suppresses the magnetism monotonically with no signs of superconductivity up to $x = 0.6$ [10]. The non-isoelectronic Cd substitution [8] at the In site enhances the AF interactions in $Ce_2RhIn_8$. A recent ARPES study [11] on a $Ce_2RhIn_{7.79}Cd_{0.21}$ sample has revealed suppression of the hybridization gap by Cd substitution. On the other hand, adding one extra electron by substitution of non-isoelectronic Sn suppresses the magnetism gradually [12]. Studies of substitutions for Ce have been focused on La [13] and Pr [14] as substituents. Both types of substitutions suppress the magnetic-ordering temperature due to the dilution effect.

In contrast to $Ce_2RhIn_8$, $Ce_2PdIn_8$ is an ambient-pressure HF superconductor with a sample dependent SC transition temperature with the highest value of $T_c \sim 0.7$ K [15-17]. No indications of magnetic ordering have been found. The compound has been linked to $CeCoIn_5$ because of its ambient-pressure superconductivity and non-Fermi-liquid behavior [18]. Because the growth of pure $Ce_2PdIn_8$ single crystals has turned out to be very difficult, only a very limited number of measurements performed on pure single crystalline samples have been reported [3,15] and no attempt of tuning of the quantum criticality by substitution has been made. According to a recent study [19] of the related $CeRh_{1-x}Pd_xIn_5$ system, the AF state of $CeRh_{1-x}Pd_xIn_5$ is almost independent of the Pd content (up to $x = 0.25$) while the pressure-induced superconductivity is enhanced. The results suggest that at higher Pd concentration the ground state of $CeRhIn_5$ might turn into a SC state already at ambient pressure. The aim of the present work is to study the evolution of antiferromagnetism and superconductivity in $Ce_2Rh_{1-x}Pd_xIn_8$ compounds with varying composition of the 4$d$-transition-metal sublattice. We report on results of single-crystal growth of selected Rh-Pd concentrations, characterization of the chemical composition and crystal structure, measurements of the temperature dependence of the specific heat in various magnetic fields and magnetization measurements as a function of temperature and magnetic field.

## 2. Sample synthesis and characterization

The $Ce_2Rh_{1-x}Pd_xIn_8$ single crystals have been grown from In self-flux by a technology procedure analogous to the one used for $Ce_2PdIn_8$ [3,15] and $CeRh_{1-x}Pd_xIn_5$ [19]. The growth composition Ce:$Rh_{1-y}Pd_y$:In = 2:1:30 was chosen for samples with lower Pd content $y \leq 0.5$; for the samples with $y \geq 0.5$, the stoichiometry 1:3:45-55 was used as it improved the homogeneity and avoided the growth of undesired phases such as $CeIn_3$ similarly to the growth procedure of $Ce_2PdIn_8$ described in literature [15]. The batches were heated up to 950 °C, kept at this temperature for 10 hours to let the mixture homogenize properly and then cooled down slowly (~3 °C/h) to 500 °C, where the remaining flux was decanted.

The samples were checked carefully by microprobe analysis using a Scanning Electron Microscope (SEM) Tescan Mira LMH equipped with an energy dispersive x-ray detector (EDX) Bruker AXS and ESPRIT software. The issue regarding the determination of the real Rh-Pd contents by microprobe analysis has been thoroughly discussed in the case of $CeRh_{1-x}Pd_xIn_5$ samples [19]. Representative crystals with final concentrations $x = 0, 0.1, 0.15, 0.28, 0.45, 0.55, 0.85$ and 1 have been selected. No crystals with Pd concentration between 0.55 and 0.85 have been found in any growth batch. In this concentration region, we have not succeeded in preparing homogeneous single crystals free of spurious phases. Similar to $CeRh_{1-x}Pd_xIn_5$ crystals [19], the actual Pd content in the samples has been found to be lower than the nominal Pd concentration of the melt. Within the crystals selected for bulk measurements, deviations from the composition $\Delta x = 0.02$-$0.05$ have been observed. For crystals with $0.3 < x < 1$ the deviation was somewhat larger being $\Delta x = 0.11$. The single crystals had a plate-like shape with the *c*-axis perpendicular to the largest surface. The mass of the samples selected for measurements varied from 0.3 to 4 mg.

The lattice parameters have been determined by single-crystal x-ray diffraction using Rigaku RAPID II. To confirm the crystal orientation, the Laue method was used. The specific heat was measured in the temperature range 0.35-300 K using PPMS 9T (*Quantum Design*) equipped with a Helium-3 option for low-temperature operation. The magnetization was measured in the temperature range 1.8-300 K using a MPMS 7 T.

## 3. Results and discussion

Single-crystal x-ray diffraction confirmed that the $Ce_2Rh_{1-x}Pd_xIn_8$ crystals adopt the $Ho_2CoGa_8$-structure type for all prepared concentrations $x = 0.10, 0.15, 0.30, 0.45, 0.55$ and $0.85$. With increasing *x*, the room-temperature lattice parameter *a* increases slightly while *c* decreases (see Fig. 1). As a result, the unit-cell volume increases with increasing Pd content whereas the *c*/*a* ratio decreases by about 1 %. The change of the *c*/*a* ratio is similar to corresponding changes in Ir- (0.9 %) and Co-substituted (0.6 %) samples [7,10] in the whole concentration range.

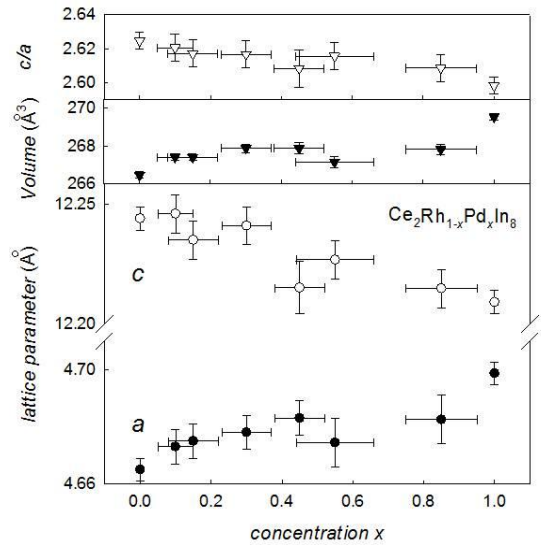

**Fig. 1.** Lattice parameters *a* and *c*, *c*/*a* ratio and unit-cell volume of $Ce_2Rh_{1-x}Pd_xIn_8$ compounds. The data for $Ce_2PdIn_8$ have been taken from literature [15].

In Fig. 2, the specific heat of $Ce_2Rh_{1-x}Pd_xIn_8$ for various Pd concentrations is shown in the $C_p/T$ vs. *T* representation. For $x = 0$, the pronounced peak at $T_N = 2.85$ K reflects the well-known transition of $Ce_2RhIn_8$ to commensurate AF ordering [2]. At lower temperatures, additional anomaly at $T_{LN} = 1.45$ K is observed which can be ascribed to a transition from the commensurate magnetic order existing between $T_{LN}$ and $T_N$, to the ground-state incommensurate order reported in literature [4]. This order-to-order transition is absent in samples with $x \geq 0.1$. The Néel temperature gradually decreases with increasing Pd concentration. For the $x = 0.45$ sample, no magnetic phase transition is observed at temperatures accessible with our experimental setup ($\geq 0.35$ K). As $T_N$ is suppressed, a pronounced upturn of the $C/T$ curve appears at low temperatures for the $x = 0.55$ sample which might be a sign of critical quantum AF fluctuations [20] (see inset in Fig. 2). For the $x = 0.85$ sample, the low-temperature upturn is already hardly visible and it is completely suppressed for $Ce_2PdIn_8$. In the $Ce_2PdIn_8$ crystal, superconductivity is observed below $T_c = 0.55$ K, in agreement with literature results discussed in Ref. 15. The value of $T_c$ of $Ce_2PdIn_8$ samples may vary between 0.5 and 0.7 K, but none of the $Ce_2Rh_{1-x}Pd_xIn_8$ crystals, which contain Rh, shows any sign of a transition to the SC state at temperatures down to 0.35 K.

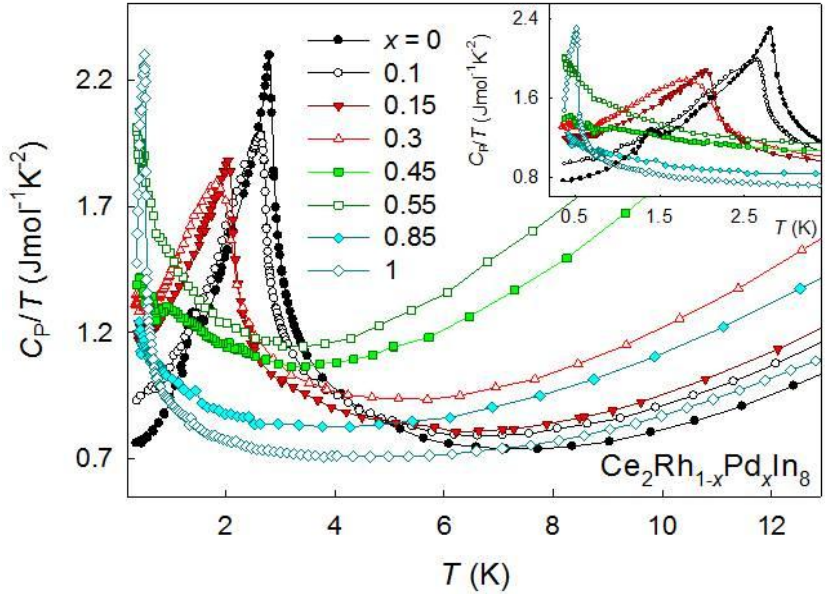

**Fig. 2.** Temperature dependence of the specific heat (plotted as $C_p/T$ vs $T$) of the $Ce_2Rh_{1-x}Pd_xIn_8$ compounds for various Pd concentrations in zero magnetic field. The inset depicts the low-temperature region.

The values of the Sommerfeld coefficient $\gamma$, which have been determined from linear fits of $C/T = \gamma + \beta T^2$ dependences in the interval 8 K < $T$ < 13 K, are shown in the phase diagram in Fig. 5. The value of $\gamma$ monotonously increases with increasing Pd content from 240 mJ mol$^{-1}$(Ce)K$^{-2}$ for $Ce_2RhIn_8$ up to the maximum value of 490 mJ mol$^{-1}$(Ce)K$^{-2}$ for $Ce_2Rh_{0.45}Pd_{0.55}In_8$, which is close to the critical Pd concentration for antiferromagnetism (the evolution of $T_N$ and $\gamma$ with increasing $x$ is shown in Fig. 5). For the sample with $x = 0.85$, $\gamma$ is dramatically reduced and it further decreases for $Ce_2PdIn_8$, exhibiting a value comparable with the $\gamma$ value for $Ce_2RhIn_8$. To allow an estimate of the magnetic entropy, the phonon term $\beta T^2$ was subtracted from the $C/T$ data. At $T_N$, a 4$f$ contribution of about 0.38 $R$ ln2 to the entropy $S_{4f}$ of $Ce_2RhIn_8$ has been determined and, for $Ce_2Rh_{0.7}Pd_{0.3}In_8$, a contribution of about 0.24 $R$ ln2. Kondo screening of the ordered Ce moments [21] probably causes that only a small fraction of the magnetic entropy is released. The influence of the Pd content on the Kondo temperature $T_K$ within the substitution series has been analyzed in more detail. There are several methods of estimating the Kondo temperature and they may lead to different values of $T_K$. However, usually a satisfactory agreement is achieved in general trends of the $T_K$ evolution within the series [4,22]. Being aware of these constraints in the present study, the value of $T_K$ was derived from the specific heat and from magnetic-susceptibility data. Using specific-heat data, the evaluation of $T_K$ was interpreted in terms of a simple two-level model considering the reduction of magnetic entropy at $T_N$. This approach considers the expression

$$\frac{\Delta S}{R} = \ln\left(1 + \exp\left(\frac{-T_K}{T_C}\right)\right) + \frac{T_K}{T_C}\left(\frac{\exp(-T_K/T_C)}{1 + \exp(-T_K/T_C)}\right) \quad (1)$$

which has been applied to the ferromagnetic CeNi$_x$Pt$_{1-x}$ series in the work of Blanco *et al.* [22]. For Ce$_2$RhIn$_8$, we have obtained $T_K \approx 6$ K which is of the same order of magnitude as $T_K \approx 10$ K reported by Malinowski *et al.* [4]. Further increase of the Kondo temperature of $T_K \approx 7.5$ K for the Pd concentration $x = 0.15$ and $T_K \approx 12$ K for $x = 0.30$ was observed simultaneously with an increase of the coefficient $\gamma$ and a decrease of the Néel temperature.

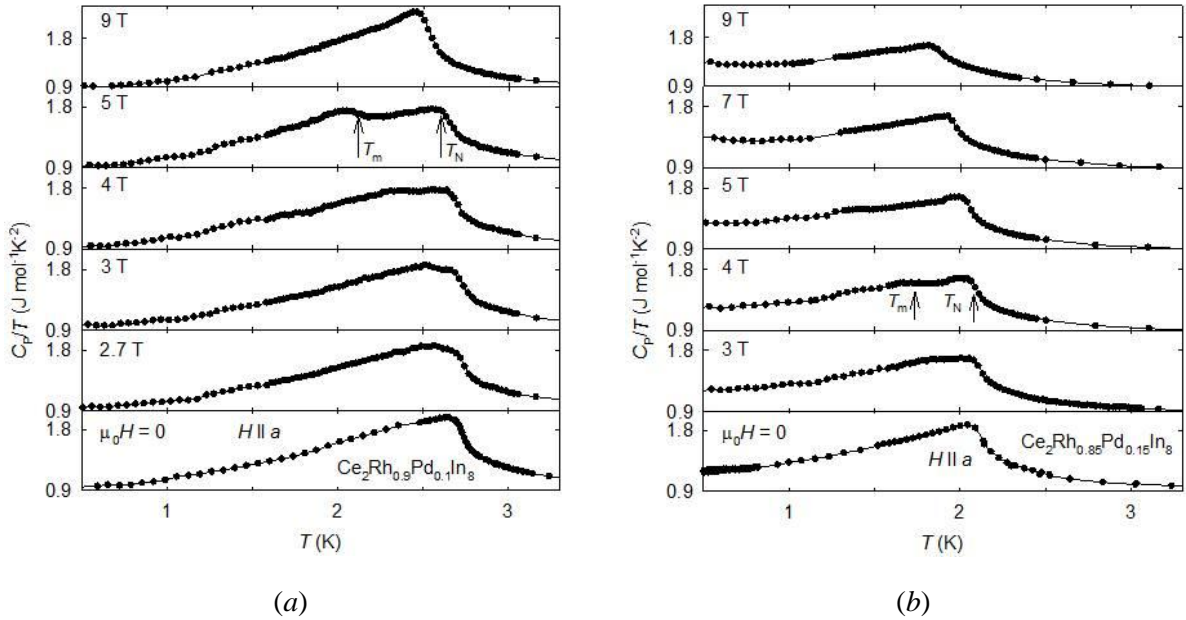

**Fig. 3.** Low-temperature part of the specific heat of (*a*) Ce$_2$Rh$_{0.9}$Pd$_{0.1}$In$_8$ and (*b*) Ce$_2$Rh$_{0.85}$Pd$_{0.15}$In$_8$, measured at various magnetic fields applied along the *a*-axis. The arrows indicate the transition temperatures $T_N$ and $T_m$.

The temperature dependence of the specific heat of Ce$_2$Rh$_{1-x}$Pd$_x$In$_8$ compounds with $x = 0.1$, 0.15 in various magnetic fields is shown in Fig. 3. As the evolution of magnetism is only modest for the magnetic field applied along the *c*-axis [23], we studied the effect of magnetic field applied within the basal plane, along the *a*-axis. A magnetic field-induced transition shows up at $T_m \approx 2.3$ K ($T_m \approx 1.7$ K) in Ce$_2$Rh$_{0.9}$Pd$_{0.1}$In$_8$ (Ce$_2$Rh$_{0.85}$Pd$_{0.15}$In$_8$) in a field of about 4 T and moves to lower temperatures when the field is increased. For samples with $x > 0.15$ (not shown), the magnetic field-induced transitions are completely suppressed and only the AF transition remains. Both transition temperatures $T_m$ and $T_N$ shift to lower temperatures with increasing Pd concentration, similar to the CeRh$_{1-x}$Pd$_x$In$_5$ system [19]. The sensitivity of $T_m$ to the Pd concentration is stronger in the Ce$_2$Rh$_{1-x}$Pd$_x$In$_8$ system in line with the behavior of $T_N$. The magnetic-field-induced transitions at $T_1$ and $T_2$ observed in Ce$_2$RhIn$_8$ [23] are smeared out in the Pd-substituted samples. Thus, the magnetic phase below $T_m$ cannot be identified directly with one of the induced phases in the *T*-*H* phase diagram (same notation as in Ref. 23 is used). One can speculate that the phase II is suppressed by the increasing Pd substitution; however, no microscopic study of magnetic structure was performed to support this scenario.

In order to study the magnetocrystalline anisotropy and the Kondo screening, the magnetic susceptibility $M(T)/H$ of $Ce_2Rh_{1-x}Pd_xIn_8$ ($x$ = 0, 0.15, 0.85, 1) was measured up to room temperature in magnetic fields applied along both the principal crystallographic directions. One can immediately recognize similarities between the low-temperature behavior of the magnetic susceptibility of $Ce_2PdIn_8$ and $Ce_2Rh_{0.15}Pd_{0.85}In_8$, while the susceptibility of $Ce_2Rh_{0.85}Pd_{0.15}In_8$ resembles more the behavior of $Ce_2RhIn_8$, as expected (see Fig. 4). At temperatures above 50 K (100 K), the $a$-axis ($c$-axis) reciprocal susceptibility obeys the Curie-Weiss law (see inset of Fig. 4); the linear fit provides the values for the paramagnetic Curie temperature $\theta_P$ and the effective magnetic moment $\mu_{eff}$ listed in Table 1. The magnetic moments $\mu_{eff}$ of the substituted samples are somewhat smaller than the Hund's rule value of 2.54 $\mu_B$ for a free $Ce^{3+}$ ion. Together with the pronounced deviation of $1/\chi(T)$ from linear behavior at lower temperatures, this can be ascribed to an interplay of Kondo and crystal-field interactions, similar to the crystal-field effects reported for $CeRhIn_5$ [24] and $CeRh_{1-x}Pd_xIn_5$ [19].

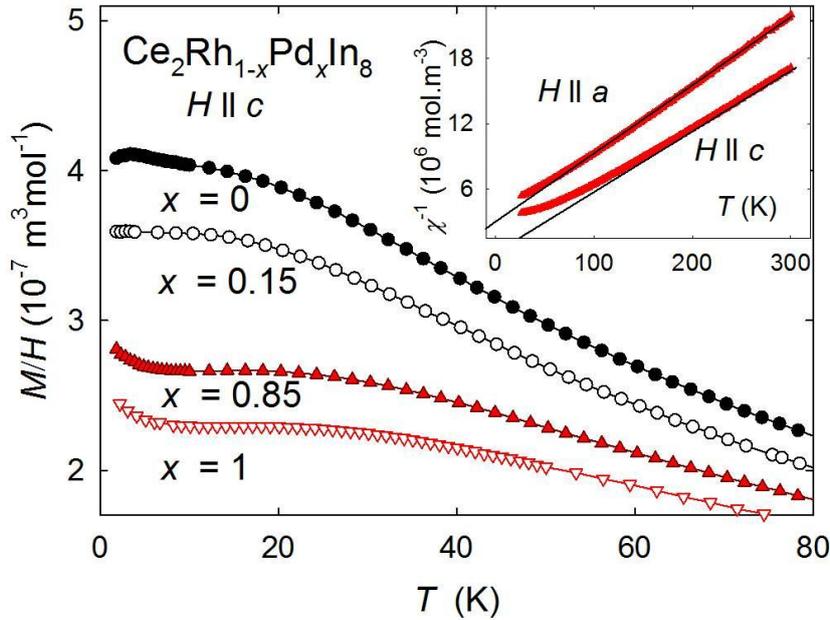

**Fig. 4.** Temperature dependence of the magnetic susceptibility $M/H$ of $Ce_2Rh_{1-x}Pd_xIn_8$ ($x$ = 0, 0.15, 0.85 1) measured in a magnetic field of 5 T applied along the $c$-axis. The inset shows the reciprocal magnetic susceptibility of $Ce_2Rh_{0.15}Pd_{0.85}In_8$ in the whole temperature range up to 300 K in magnetic fields applied along the $a$- and $c$-axis. The lines represent the fits to the Curie-Weiss law.

According to Hewson [25], the order of magnitude of the Kondo temperature $T_K$ can be estimated from the paramagnetic Curie temperature using the formula $|\theta_P|/4$. One can immediately see a significant difference of the $\theta_P$ value for $Ce_2PdIn_8$ and the substituted samples likely reflecting the unique SC ground state of $Ce_2PdIn_8$ within the series. Considering the concentration evolution of $\theta_P^a$, $T_K$ decreases first from $T_K$ = 17 K for $Ce_2RhIn_8$ to 12 K for $Ce_2Rh_{0.15}Pd_{0.85}In_8$ and it is increases again for $Ce_2PdIn_8$ ($T_K$ = 23 K). The concentration evolution of $\theta_P^c$ is non-monotonous with a weak tendency to increase from $T_K \sim 0$ K to 13. Considering the polycrystalline average $(2 \cdot \theta_P^a + \theta_P^c)/3$ of both sets of values, the Kondo temperature of the substituted samples decreases moderately (from 12 K to 9 K) which is in conflict with the $T_K$ evolution derived from the specific heat. According to a general rule [22], the 4$f$-ligand hybridization weakens with increasing cell volume. Taking into account the probable weakening of the hybridization strength when Pd is substituted instead of Rh, one might expect naively that $T_K$ decreases with increasing Pd concentration instead of its enhancement as proposed by the two-level model above; thus the model seems to be inappropriate in the case of the Ce$_2$Rh$_{1-x}$Pd$_x$In$_8$ series. The increasing cell volume would also suggest an increase of the robustness of the AF order. However, $T_N$ decreases with the Pd content which points to a combined effect of more energy scales playing important role in the series including the exchange interaction.

**Table 1**

Concentration dependence of the paramagnetic Curie temperature $\theta_P$ and the effective magnetic moment $\mu_{eff}$ of Ce$_2$Rh$_{1-x}$Pd$_x$In$_8$ ($x = 0, 0.15, 0.85, 1$).

|      | $\theta_P$ (K) |          | $\mu_{eff}$ ($\mu_B$/Ce) |          |
| ---- | -------------- | -------- | ------------------------ | -------- |
| $x$  | $\parallel a$  | $\parallel c$ | $\parallel a$       | $\parallel c$ |
| 0    | -68            | -8       | 2.5                      | 2.6      |
| 0.15 | -60            | -3       | 2.4                      | 2.4      |
| 0.85 | -48            | -17      | 2.3                      | 2.4      |
| 1    | -90            | -50      | 2.6                      | 2.6      |

The concentration dependence of the magnetic-ordering temperature $T_N$ and the electronic Sommerfeld coefficient are depicted in the temperature-concentration diagram in Fig. 5. For $T_{LN}$ for Ce$_2$RhIn$_8$ and $T_c$ for Ce$_2$PdIn$_8$, only one single point is shown because no signs of these transitions have been found for the neighboring compositions Ce$_2$Rh$_{0.9}$Pd$_{0.1}$In$_8$ and Ce$_2$Rh$_{0.15}$Pd$_{0.85}$In$_8$. Apparently, no overlap of superconductivity and antiferromagnetism is observed at temperatures above 0.35 K, i.e. either the phenomena do not coexist and they are well separated in a specific concentration range or the coexistence is hidden at temperatures below 0.35 K.

The critical concentration where $T_N$ goes to zero can be extrapolated to $x \sim 0.45$. The low-temperature upturn of the $C/T$ curve observed in this concentration region suggests presence of quantum fluctuations [20]. As the $C/T$ upturn reduces before $x$ approaches 1, the quantum fluctuations are likely not related to the superconductivity of Ce$_2$PdIn$_8$.

Similar to the Co [10] and Ir [7] substitutions, Pd substitution in Ce$_2$RhIn$_8$ suppresses gradually the AF order. Nevertheless, the magnetism of Ce$_2$Rh$_{1-x}$T$_x$In$_8$ is more sensitive to the Pd concentration than to the isoelectronic substitutions ($T$ = Co, Ir) because $T_N$ is suppressed rapidly at low Pd content. The Ce$_2$Rh$_{1-x}$Co$_x$In$_8$ and Ce$_2$Rh$_{1-x}$Pd$_x$In$_8$ systems do not reveal any SC transition for magnetically ordered samples. As Ce$_2$CoIn$_8$ ($T_c = 0.4$ K [26]) and Ce$_2$PdIn$_8$ are HF superconductors, one might recognize similarities of the corresponding $T$-$x$ phase diagrams. According to Ref. 7, the occurrence of the SC phases becomes unfavorable for alloys with $n = 2$ due to the higher dimensionality and stronger disorder. This scenario is supported by the fact that the SC state is suppressed in Ce$_2$PdIn$_8$ immediately after introducing Rh substitution. Unfortunately, the phase diagram of Ce$_2$Rh$_{1-x}$Co$_x$In$_8$ is not known for $0.6 < x < 1$ which prevents us from any further speculations.

Comparison of $Ce_2Rh_{1-x}Pd_xIn_8$ and the related $CeRh_{1-x}Pd_xIn_5$ alloys [19] shows that the suppression of magnetic order by moderate concentrations of Pd ($x < 0.3$) is stronger in the systems with $n = 2$. Similar behavior is observed for compounds substituted with Co and Ir compared to their more 2D counterparts ($n = 1$) [9,27].

The character of the phase diagram (Fig. 5) resembles recently published results for the tetragonal $CeRh(In,Sn)_5$ [28] and $U_2Pt_xRh_{1-x}C_2$ [20] systems. In the Sn-substituted $CeRhIn_5$, the magnetic order is suppressed at a critical concentration $x_c \sim 0.35$ and no superconductivity appears around this concentration. In $U_2Pt_xRh_{1-x}C_2$, the antiferromagnetism is suppressed completely at $x \sim 0.7$ whereas the superconductivity emerges above $x \sim 0.9$; thus, no coexistence of both phenomena is present in these systems. Similar to the present results, a low-temperature upturn of the specific heat is observed for the critical concentrations; in the case of $Ce_2Rh_{1-x}Pd_xIn_8$, it is accompanied with a simultaneous increase of the $\gamma$ coefficient.

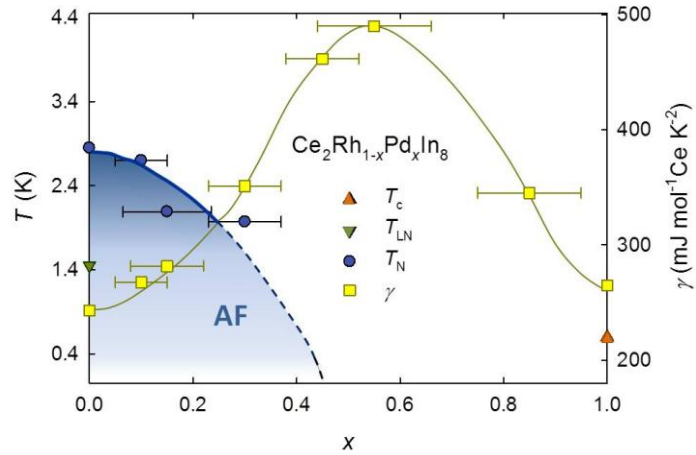

**Fig. 5.** $T$-$x$ phase diagram of $Ce_2Rh_{1-x}Pd_xIn_8$ based on the results of specific-heat measurements. The temperatures $T_N$, $T_{LN}$ and $T_c$ are presented by the left axis; the Sommerfeld coefficient $\gamma$ by the right axis. The solid lines are guides to the eyes; the dashed line is an extrapolation of the data to the zero value of $T_N$: the critical concentration is $x \sim 0.45$. The error bars have been obtained from the microprobe analysis.

## 4. Conclusions

$Ce_2Rh_{1-x}Pd_xIn_8$ ($x = 0, 0.1, 0.15, 0.3, 0.45, 0.55, 0.85, 1$) single crystals have been successfully grown by means of the In self-flux method. A series of specific-heat and magnetization measurements has been performed in order to study the effect of Pd substitution on the ground-state properties of $Ce_2RhIn_8$. The measurements show that the Néel temperature $T_N$ is gradually suppressed upon Pd substitution and, simultaneously, the Sommerfeld coefficient $\gamma$ is enhanced. The AF order vanishes at a critical concentration $x \sim 0.45$; the maximum of the $\gamma(x)$ curve corresponds to this critical concentration region where no magnetic order is observed and, moreover, the emergence of critical fluctuations reflected in the low-temperature specific-heat upturn below $T = 1$ K coincides with this region. The order-to-order transition at $T_{LN}$ and the magnetic field-induced transition $T_m$ are also affected by the introduction of: increasing the Pd concentration suppresses $T_{LN}$ immediately and $T_m$ is not observable for Pd contents higher than $x > 0.15$. Ambient pressure superconductivity is not observed in any of the Rh-containing samples.

In general, the Ce$_2$Rh$_{1-x}$Pd$_x$In$_8$ single crystals have been found more sensitive to Pd substitution compared to the isoelectronic substitutions with Co and Ir [7,10] and also compared to their 2D CeRh$_{1-x}$Pd$_x$In$_5$ analogs [19]. The *T-x* phase diagram suggests that magnetic order and superconductivity do not coexist in this system and shows that both phenomena are well separated in the concentration range $0.3 < x < 1$. Nevertheless, the slight possibility that coexistence of magnetic order and superconductivity may be hidden at temperatures which are not accessible with our experimental equipment cannot be entirely excluded. Further low-temperature studies are desired to resolve this still open question.

**Acknowledgements**

This work was supported by the Grant Agency of Charles University (Project no. 134214) and by the Czech Science Foundation (Project no. GP14-17102P). Experiments were performed at MLTL (http://mltl.eu/), which is supported within the program of Czech Research Infrastructures (Project no. LM2011025).